\documentclass[11pt,a4paper]{article}
\pdfoutput=1

\usepackage{jheppub}

\usepackage{amsmath,amsthm,amsfonts,amssymb,citesort,amscd,mathrsfs,graphicx,fontenc,Glebdefs,
}

\usepackage{color}
\let\normalcolor\relax
\usepackage{colordefs}

\newcommand{\XX}{\Theta}
\newcommand{\YY}{\Psi}
\newcommand{\driv}{\mathscr{D}}
\newcommand{\YRw}[1]{Y^{(R)}_{#1|w}}
\newcommand{\YLw}[1]{Y^{(L)}_{#1|w}}

\newcommand{\YRm}{Y^{(R)}_{-}}
\newcommand{\YLm}{Y^{(L)}_{-}}
\newcommand{\YRpm}{Y^{(R)}_{\pm}}
\newcommand{\YLpm}{Y^{(L)}_{\pm}}
\newcommand{\YRvw}[1]{Y^{(R)}_{#1|vw}}
\newcommand{\YLvw}[1]{Y^{(L)}_{#1|vw}}
\newcommand{\RwXX}[1]{\varrho_{\scriptscriptstyle #1}}
\newcommand{\RvwXX}[1]{r_{\scriptscriptstyle #1}}
\newcommand{\RwYY}[1]{\rho_{\scriptscriptstyle #1}}
\newcommand{\RvwYY}[1]{{\sf r}_{\scriptscriptstyle #1}}

\title{Lifting asymptotic degeneracies with the Mirror TBA}

\author[a]{Alessandro Sfondrini}
\author[a]{and Stijn J. van Tongeren}
\affiliation[a]{Institute for Theoretical
Physics and Spinoza Institute,\\Utrecht University, 3508 TD
Utrecht, The Netherlands}

\emailAdd{A.Sfondrini@uu.nl}
\emailAdd{s.j.vantongeren@uu.nl}

\abstract{We describe a qualitative feature of the $\AdS$ string spectrum which is not captured by the asymptotic Bethe ansatz. This is reflected by an enhanced discrete symmetry in the asymptotic limit, whereby extra energy degeneracy enters the spectrum. We discuss how finite size corrections should lift this degeneracy, through both perturbative (L\"uscher) and non-perturbative approaches (the Mirror TBA), and illustrate this explicitly on two such asymptotically degenerate states.}

\begin{document}
\renewcommand{\thefootnote}{\arabic{footnote}}
\setcounter{footnote}{0}

\begin{flushright}\small{ITP-UU-11/21\\SPIN-11/16}\end{flushright}

\maketitle

\section{Introduction}

In the decompactification limit, both the light-cone gauge fixed $\AdS$ superstring and its AdS/CFT dual \cite{Maldacena:1997re} $\mathcal{N}=4$ super Yang-Mills theory have a description through an asymptotic Bethe ansatz\footnote{For a review of integrability in the AdS/CFT correspondence see \cite{Arutyunov:2009ga,Beisert:2010jr}.}. This description does not apply to either theory at finite size, where the only current non-perturbative description is through a set of equations known as the mirror thermodynamic Bethe ansatz (TBA) equations for the superstring.

The idea of applying methods from integrable relativistic models at finite size \cite{Zamolodchikov:1989cf} to the AdS/CFT correspondence was initiated in \cite{Ambjorn:2005wa} and explored in detail in \cite{Arutyunov:2007tc}. The main step in deriving the mirror TBA equations is the formulation of the string hypothesis \cite{Takahashi:19721dHubbard}, which was done for the present model in \cite{Arutyunov:2009zu} by using the mirror version of the Bethe-Yang equations \cite{Beisert:2005fw} for the $\ads$ superstring. This was followed by a derivation of the canonical \cite{Arutyunov:2009ur,Bombardelli:2009ns,Gromov:2009bc} and simplified \cite{Arutyunov:2009ux} TBA equations that describe the ground state of the theory\footnote{The associated Y-system was conjectured in \cite{Gromov:2009tv}.}. These equations have been used to analyze the vanishing of the ground state energy of the theory at finite size \cite{Frolov:2009in}. Importantly, these ground state equations can be used to obtain equations for the excited states, through a contour deformation trick \cite{Arutyunov:2011uz} inspired by the analytic continuation procedure of \cite{Dorey:1996re}. Using the contour deformation trick, the mirror TBA equations have been used to reproduce perturbative results found through L\"uscher's approach\footnote{The use of L\"uscher's approach \cite{Luscher:1985dn} in the AdS/CFT correspondence was first advocated in \cite{Ambjorn:2005wa}.} \cite{Bajnok:2009vm,Arutyunov:2010gb,Balog:2010xa}, and to study certain
states in the $\alg{sl}(2)$ sector in considerable detail \cite{Gromov:2009tq,Arutyunov:2009ax,Balog:2010vf}, specifically at intermediate coupling in \cite{Gromov:2009zb,Frolov:2010wt}. The analytic properties of the Y-functions \cite{Arutyunov:2009ax,Cavaglia:2010nm,Cavaglia:2011kd,Arutyunov:2011inprogress} are essential in determining these equations, and have proved useful for further understanding of the mirror TBA equations. Following these developments, discontinuity relations can now be used to find the excited state equations directly in the $\alg{sl}(2)$ subsector \cite{Balog:2011nm}, giving results in complete equivalence with the contour deformation trick where applicable. Moreover, the simplified TBA equations have recently been brought to a quasi-local form \cite{Balog:2011cx}, another step in the direction of obtaining a so-called non-linear integral equation (NLIE) description of the spectral problem at finite size. Additional steps in this direction had already been taken in \cite{Gromov:2010km,Suzuki:2011dj}.

It is well known that the asymptotic Bethe ansatz (ABA) captures the leading $1/J$ corrections to the asymptotic energy spectrum, while it misses the exponential corrections due to the finite system size. Therefore the ABA misses quantitative information on the spectrum, as is for example clearly illustrated by wrapping corrections to scaling dimensions in the gauge theory. Following an observation of \cite{Arutyunov:2011uz}, in this paper we will show that for the $\AdS$ superstring finite size effects are not only quantitative in nature. In fact we will demonstrate that the ABA also misses \emph{qualitative} information on the spectrum, owing to a discrete symmetry enhancement of the model in the asymptotic limit, so that certain states become degenerate asymptotically.

This paper is organized as follows. In the next section we start by discussing the symmetries and degeneracies of the asymptotic Bethe ansatz and explain the asymptotic symmetry enhancement alluded to just above. Also, we will indicate how finite size effects should lift this degeneracy. After painting the general picture we illustrate these ideas by considering two concrete states with degenerate energies in the asymptotic limit. We will show that these states have manifestly different TBA equations and explicitly compute the different finite size corrections they receive, in line with the general discussion.

\section{Extra degeneracy in the asymptotic limit}

In general the energy spectrum of string states is expected to have degeneracies, owing to the superconformal symmetry of the model. Because of this symmetry, string states arrange themselves in superconformal multiplets which each share a common energy. At the level of the (asymptotic) Bethe ansatz these degeneracies are reflected by the fact that solutions to the Bethe-Yang equations only give the highest weight states of the underlying symmetry algebra, familiar from e.g. the Heisenberg \cite{Faddeev:1996iy} and Hubbard model \cite{Essler:1991wg}. Completeness of the Bethe ansatz then follows by adding the states which lie in the same multiplet, which then by construction have the same energy. In the case of the asymptotic Bethe ansatz for the $\AdS$ superstring however, there is \emph{additional} degeneracy, degeneracy which arises in the decompactification limit and should not be present in the complete model. This degeneracy occurs in the asymptotic Bethe ansatz due to enhanced symmetry in the asymptotic limit, indicating qualitative features of the model that are not captured by the asymptotic solution.

In the light-cone gauge the superstring has manifest $\su_L(2|2)\oplus \su_R(2|2)$ symmetry, where the subscript $L$ and $R$ distinguish the two $\su(2|2)$ factors, conventionally called left and right. By construction, the model possesses a $\mathbb{Z}_2$ symmetry, which we call left-right symmetry, interchanging the sets of left and right $\su(2|2)$ charges. This means that for every state with a given set of $\su_L(2|2)\oplus \su_R(2|2)$ charges, there exists a state with equal energy, with left-right interchanged $\su(2|2)$ charges. It is this left-right symmetry which is enhanced in the asymptotic limit to a larger discrete group.
At the level of the Bethe-Yang equations this enhancement is manifested by the fact that they allow more than just an interchange of complete sets of left and right charges, actually allowing free redistribution of roots between the left and right sectors in certain cases. In the finite size model on the contrary, there is no reason to assume states related by such a redistribution should have the same energy, so we expect that finite size effects lift this asymptotic degeneracy. We will now consider these ideas in more detail.

\subsection{The Bethe-Yang equations}

The Bethe-Yang equation for the $\AdS$ superstring in the light-cone gauge is given by \cite{Beisert:2005fw}
\begin{equation}\label{equ:betheyangmain}
1= e^{ip_k J} \prod_{\textstyle\atopfrac{l=1}{l\neq
k}}^{K^{\mathrm{I}}}S_{\sl(2)}(p_k,p_l)\prod_{\alpha = L,R} \prod_{l=1}^{K^{\mathrm{II}}_{(\alpha)}} \frac{x^-_k-y_l^{(\alpha)}}{x^+_k-y_l^{(\alpha)}}\sqrt{\frac{x^+_k}{x^-_k}}\,,\ \ \ \ \ \ \ \ k=1,\dots K^{\rm I}\;.
\end{equation}
In addition to the rapidities of fundamental particles, this equation contains $y^{(\a)}$ roots. Together with the $w^{(\a)}$ roots ($\a=L,R$) which enter in the auxiliary Bethe equations just below, these correspond to the $\su_L(2|2)\oplus\su_R(2|2)$ symmetry of the model. The auxiliary Bethe equations consist of two independent sets of two coupled equations for the $y$ and $w$ roots, given by
\begin{align}
\label{equ:betheyangaux1}
1=&\prod_{l=1}^{K^{\mathrm{I}}}\frac{y_{k}^{(\a)}-x^{-}_{l}}{y_{k}^{(\a)}-x^{+}_{l}}\sqrt{\frac{x_l^+}{x_l^-}}
\prod_{l=1}^{K^{\mathrm{III}}_{(\a)}}\frac{\nu_{k}^{(\a)}-w_{l}^{(\a)}+\frac{i}{g}}{\nu_{k}^{(\a)}-w_{l}^{(\a)}-\frac{i}{g}}\;, & k = 1, \ldots, K^{\mathrm{II}}_{(\a)}, \; \, &\a=L,R\;,\\
\label{equ:betheyangaux2}
1=&\prod_{l=1}^{K^{\mathrm{II}}_{(\a)}}\frac{w_{k}^{(\a)}-\nu_{l}^{(\a)}+\frac{i}{g}}{w_{k}^{(\a)}-\nu_{l}^{(\a)}-\frac{i}{g}}
\prod_ {\textstyle\atopfrac{l=1}{l\neq
k}}^{K^{\mathrm{III}}_{(\a)}}\frac{w_{k}^{(\a)}-w_{l}^{(\a)}-\frac{2i}{g}}{w_{k}^{(\a)}-w_{l}^{(\a)}+\frac{2i}{g}}\;, & k = 1, \ldots, K^{\mathrm{III}}_{(\a)}, \; \, &\a=L,R\;,
\end{align}
where $\nu_{k}^{(\a)} = y_{k}^{(\a)} + 1/y_{k}^{(\a)}$. The fact that we have two sets of identical left-right decoupled equations corresponds directly to the left-right symmetry mentioned earlier. At the level of the transfer matrix this is reflected by the fact that we have a transfer matrix for each sector, $T^{(L)}$ and $T^{(R)}$; both are $\su(2|2)$-invariant transfer matrices with eigenvalues parametrized by the auxiliary roots, and these eigenvalues are therefore arranged in $\su(2|2)$ multiplets. As mentioned, solutions of the auxiliary equations (\ref{equ:betheyangaux1}-\ref{equ:betheyangaux2}) identify highest weight states of the $\su(2)$ subalgebras of this symmetry algebra, labeled by the Dynkin labels $(s_{(\alpha)}, q_{(\alpha)})$. The weights are encoded in the excitations numbers as
\begin{equation}
s_{(\alpha)}=K^{\mathrm{I}}-K^{\mathrm{II}}_{(\a)}\,, \ \ \ \ q_{(\a)}=K^{\mathrm{II}}_{(\a)}-2K^{\mathrm{III}}_{(\a)}\;,
\end{equation}
where the excitation numbers satisfy
\begin{equation}
K^{\mathrm{I}}\geq K^{\mathrm{II}}_{(\a)}\geq 2 K^{\mathrm{III}}_{(\a)}\,,\ \ \a=L,R\;.
\end{equation}

\subsection{Extra degeneracy}

As indicated, the discrete left-right symmetry of the light-cone gauge fixed model can be enhanced in the asymptotic limit, giving a higher amount of degeneracy in the spectrum. This is the case when $K^{\mathrm{III}}_{(\a)} = 0$, for states with a given total number of $y$ roots; $\sum_\a K^{\mathrm{II}}_{(\a)} = K^{\mathrm{II}}_{\scriptscriptstyle \rm Tot} > 1$.
\emph{For any such state}, the auxiliary Bethe equations reduce to
\begin{equation}
\label{eq:}
\prod_{l=1}^{K^{\mathrm{I}}}\frac{y_{k}^{(\a)}-x^{-}_{l}}{y_{k}^{(\a)}-x^{+}_{l}}\sqrt{\frac{x_l^+}{x_l^-}} = 1\,, \; \; \; k = 1, \ldots,  K^{\mathrm{II}}_{(\a)}\,, \; \a=L,R\,,\;
\end{equation}
showing that we have \emph{one and the same equation for each of the $y_{k}^{(\a)}$} \cite{Arutyunov:2011uz}. The number of solutions we can pick for each $y$ depends on the main excitation number $K^{\mathrm{I}}$, and in general we must take care to only allow for regular configurations of roots. Nonetheless, the consequence of this degeneration is immediately clear: provided there is more than one allowed solution for $y_{k}^{(\a)}$ we can freely redistribute any number of different $y$ roots between the left and right sectors \emph{without changing the main Bethe-Yang equation}, because it itself contains a product over the left and right roots. This is the enhancement of the left-right symmetry of the finite size model in the asymptotic limit. Let us note that the corresponding symmetry group acts on regular highest weight states only, and that this action cannot be extended to the other states.

Two states differing by such a redistribution will have the same asymptotic momentum, hence energy, while there is no reason to assume their energies should be identical outside the asymptotic regime. Rather, it would actually be a surprising coincidence if their energies agreed. Stated more strongly, looking at the description of the finite size model through the mirror TBA it should be conceptually clear that this symmetry is only present asymptotically; the presence of $w$ roots generically spoils this symmetry, and while an individual state might have no $w$ roots, in the mirror TBA such a state is described in interaction with a thermal background containing all possible excitations. Note also that two such asymptotically equivalent states have manifestly different Dynkin labels, while they are not in the same superconformal multiplet. Indeed, such states correspond to potentially wildly different operators in $\mathcal{N} = 4$ SYM.

\subsection{Lifting degeneracies through finite size effects}

As just stated, we expect finite size corrections to lift this degeneracy of the asymptotic spectrum, whether it be through the thermodynamic Bethe ansatz, or perturbatively through L\"uscher corrections. How this happens is perhaps most immediately seen through the complete formula for the energy of a string state in the mirror TBA approach \cite{Arutyunov:2009ur}
\begin{equation}
\label{eq:energy}
E=\sum_{k=1}^{K^{\mathrm{I}}}\mathcal{E}_k-\frac{1}{2\pi} \sum_{Q=1}^{\infty} \int dv\,\frac{d\tilde{p}^Q}{dv}\,\log(1+Y_Q)\;,
\end{equation}
where $\mathcal{E}_k=i\tilde{p}(u_{*k})$ gives the asymptotic contribution to the energy, while the second term arises from the finite size of the system. In this formula, the $Y_Q$-functions are determined through the mirror TBA equations, which intricately couple the auxiliary left and right $Y$-functions. Now in general there is no reason to expect that the TBA equations for two of these asymptotically degenerate states should be the same, meaning they should receive different finite size corrections, lifting the degeneracy of the asymptotic spectrum.

Alternately, expanding the energy formula (\ref{eq:energy}) to leading order around the asymptotic solution (small $Y_Q$-functions) gives a formula in direct agreement with L\"uscher's approach
\begin{align}\label{eq:perturbativeenergy}
E_{LO} = -\frac{1}{2\pi}\sum_{Q=1}^{\infty}\int dv \frac{d\tilde{p}}{dv} Y^{\circ}_{Q}(v).
\end{align}
Here $Y^{\circ}_{Q}$, in the above expanded to leading order in the coupling constant, is given by the generalized L\"uscher's formula \cite{Bajnok:2008bm}
\begin{align}
\label{eq:YQasympt}
Y^{\circ}_Q(v) = e^{-J\tilde{\mathcal{E}}_Q(v)}\;T^{(L)}(v|\vec{u})\;T^{(R)}(v|\vec{u})\,\prod_k S^{Q1_*}_{\alg{sl}(2)}(v,u_k).
\end{align}
In this formula $\tilde{\mathcal{E}}_Q(v)$ is the energy of a mirror $Q$-particle, $S^{Q1_*}_{\alg{sl}(2)}(v,u_k)$ denotes the $\alg{sl}(2)$ $S$-matrix with arguments in the mirror ($v$) and string regions ($u_k$) and finally $T^{(L,R)}$ are the left and right transfer matrices, given in appendix \ref{ap:transfermatrices}. Clearly these corrections couple the left and right sectors through the product of transfer matrices. Such transfer matrices, and more importantly their products $T^{(L)}T^{(R)}$, will generically be different for two asymptotically degenerate states, giving different perturbative finite size corrections, showing again that the denegeracy of the asymptotic spectrum is lifted in the finite size theory.

In what follows we will illustrate these ideas concretely for two different four-particle states, both parametrized by two $y$ roots. We will show how the full sets of mirror TBA equations describing these states are manifestly different (though naturally in an elegant symmetric way), and explicitly compute different leading-order corrections to the energy, which hence lift the asymptotic degeneracy. Let us first introduce our states.

\section{Two explicit states}

We consider two states that have the same value for $K^{\mathrm{I}}$ and $\sum_\alpha K^{\mathrm{II}}_{(\alpha)}$, but different values for the individual $K^{\mathrm{II}}_{(\alpha)}$. When $K^{\mathrm{I}}=2$ there are no non-trivial level-matched solutions of the auxiliary equations, therefore we consider the states $\XX$ and $\YY$, as presented below in table \ref{tab:states}.

\begin{table}[!h]
\begin{center}
\begin{tabular}{| c | c | c | c | c | c |}
\hline
State & $K^{\mathrm{I}}$ &  $K_{(L)}^{\mathrm{II}}$ & $K_{(R)}^{\mathrm{II}}$& $K_{(\a)}^{\mathrm{III}}$ & Weights	 \\
\hline
$\XX$	& 4& 2& 0& 0 & $[2,J-1,0]_{(2,4)}$ \\
\hline
$\YY$	& 4& 1& 1& 0 & $[1,J-1,1]_{(3,3)}$ \\
\hline
\end{tabular}
\label{tab:states}
\caption{The two asymptotically degenerate states we consider. Note the manifestly different excitation numbers. For the readers' convenience we have also presented the Dynkin labels of the states, denoted by $[q_L,p,q_R]_{(s_L,s_R)}$.}
\end{center}
\end{table}

\noindent For either state we have four rapidities $u_i$, level-matched, and two auxiliary roots $y^{(\alpha)}_i$, either both left or one left and one right. In both cases we take the four rapidities to come in pairs: $u_1=-u_2>0$ and $u_3=-u_4>0$.

As discussed above, for both states the auxiliary equation for any $y^{(\alpha)}_i$ is the same. From (\ref{equ:betheyangaux1}), and imposing the rapidities to come in pairs, we find
\begin{equation}
1=\prod_{i=1}^4 \frac{y-x^-_i}{y-x^+_i},\ \ \ \mathrm{with}\ \ x^\pm_1=-x^\mp_2,\ \  x^\pm_3=-x^\mp_4.
\end{equation}
This admits two regular roots (in addition to $y=0,\infty$) that are opposite to each other, $y=\pm y_o$, where
\begin{equation}
y_o=\sqrt{\frac{x^-_1\,x^+_1(x^-_3 -x^+_3)+x^+_3\,x^-_3(x^-_1-x^+_1)}{x^-_1-x^+_1+x^-_3-x^+_3}}\,.
\end{equation}
Therefore we take $y^{(L)}_{1,2}=\pm y_o$ for state $\XX$, and $y^{(L)}_{1}=+ y_o$ and $y^{(R)}_{1}=- y_o$ for state $\YY$. We can now solve the two Bethe-Yang equations (\ref{equ:betheyangmain}) for $u_1$ and $u_3$ at a given value of $J$, by plugging in the auxiliary roots, recalling that the $\alg{sl}(2)$ $S$-matrix is given by
\begin{equation}
S_{\sl(2)}(u_1,u_2)=\sigma^{-2}\,\frac{x^+_1-x^-_2}{x^-_1-x^+_2}\,\frac{1-\frac{1}{x^-_1x^+_2}}{1-\frac{1}{x^+_1x^-_2}}\;,
\end{equation}
where $\sigma$ is the dressing factor.

Solving the resulting equation analytically is not feasible. Therefore, we first consider the limit $g\to0$, rescaling the rapidities such that they remain finite, $u_i\to \frac{1}{g}u^o_i$. Then the equations for $u_1^o$ and $u_3^o$ decouple, and both take the simple form
\begin{equation}
1=\left(\frac{u_k^o+i}{u_k^o-i}\right)^{J+2}\ \ \ \Longrightarrow\ \ \ u_k^o=\cot \frac{n_k\,\pi}{J+2},\ \ n_k=1,\dots J+1\;,
\end{equation}
where the sum of positive $n_k$ giving the string level of the state \cite{Arutyunov:2004vx}. In order to have a generic root configuration we focus on the case $J=4$ at string level three, where we can solve (\ref{equ:betheyangmain}) numerically for arbitrary values of $g$, requiring that at small coupling
\begin{equation}\label{equ:asymptoticrapidities}
u_1^o=-u_2^o=\sqrt{3}\,,\ \ \ \ \ u_3^o=-u_4^o=\frac{1}{\sqrt{3}}\;.
\end{equation}

\begin{figure}[!t]
\centering
     \begin{minipage}{\textwidth}
\centering
\includegraphics[width=10cm]{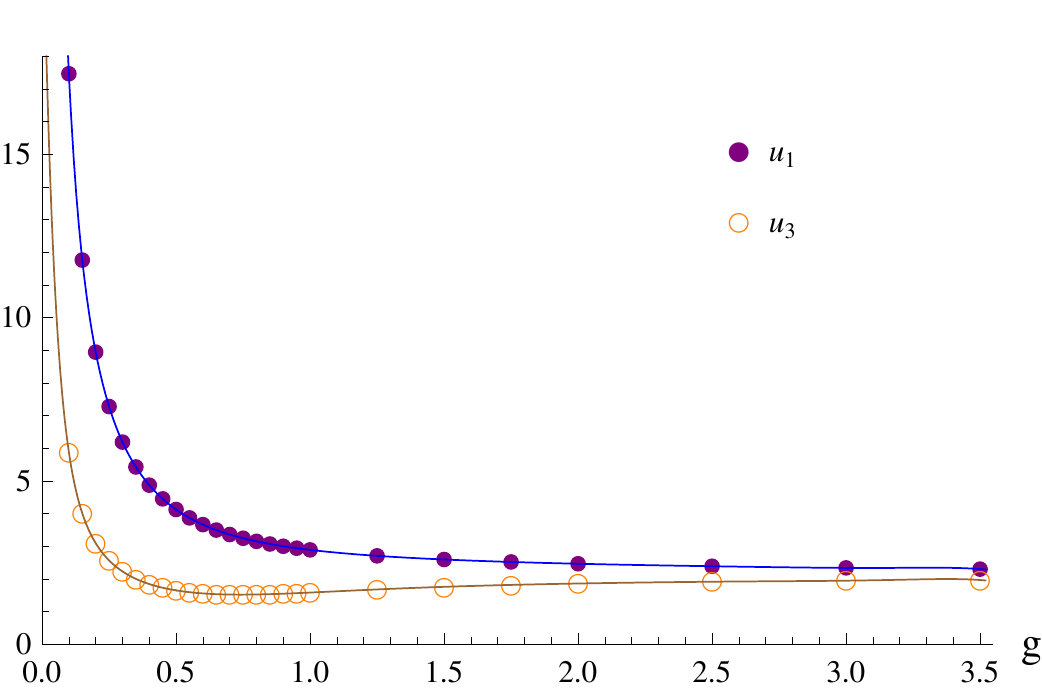}
         \caption{The rapidities $u_1$ and $u_3$ obtained from Bethe-Yang equation at different values of $g$. Note that they asymptote to two as the coupling is increased.}
     \label{fig:rapidities}
\end{minipage}
\end{figure}

In figure \ref{fig:rapidities} the numerical solutions $u_1(g)$ and $u_3(g)$ are shown. Note again that these solutions are the same for both states. In solving (\ref{equ:betheyangmain}) numerically, the representation of the dressing phase as presented in \cite{Frolov:2010wt} is most convenient. It is worth remarking that at finite values of $g$, no simple relation between $u_1$ and $u_3$ holds, despite what we see at weak coupling in (\ref{equ:asymptoticrapidities}).

Through the AdS/CFT duality, the $\XX$ state corresponds to an operator schematically of the form $\mbox{Tr}(D^2 \bar{\psi}\bar{\psi} Z^3)$. The correspondent operator of the $\YY$ state is actually a linear combination of two types of operators, namely $\mbox{Tr}(D^2 \bar{\psi} \psi Z^3 )$ and $\mbox{Tr}(D^3 W Z^4)$. In these expressions, all excitations have the highest allowed charges.

\subsection{TBA equations}

\noindent In order to obtain TBA equations for an excited state we use the contour deformation trick, following \cite{Arutyunov:2009ax}. A clear overview of this whole approach can be found in \cite{Arutyunov:2011uz}. In short we assume that the ground state and excited state TBA equations differ only by the choice of the integration contours. Upon deforming the integration contours of the excited state TBA equations to the ground state ones, we pick up additional contributions whenever there is a singularity in the physical strip of the rapidity plane. This leads to the appearance of new driving terms in the excited state TBA equations.

In the present case, we do this for the left and right sectors, for both states. We denote the $Y$-functions $\YLw{M},\YLpm,\YLvw{M}$ and $\YRw{M},\YRpm,\YRvw{M}$ for a given state in the left and right sectors respectively; the sectors are coupled by the $Y_Q$ functions. Below we discuss the analytic properties and related integration contours for both states in detail, followed by the resulting TBA equations.

\subsubsection{Analytic properties}

As discussed in  \cite{Arutyunov:2011uz}, we will use the (left and right) asymptotic $Y$-functions to study the analytic properties of the TBA equations. Their asymptotic construction is given in appendix \ref{ap:transfermatrices}. Let us stress that all the $Y$-functions but $Y_Q$ are defined in a given sector, i.e. $Y_{M|w}\equiv Y_{M|w}^{(\alpha)}$ etc. Only $Y_Q$ couples the left and right sectors, and indeed asymptotically $Y_Q$ contains the product of left and right transfer matrices, as in (\ref{eq:YQasympt}).

As shown in detail for the Konishi state in \cite{Arutyunov:2009ax}, the precise analytic structure of asymptotic $Y$-function will depend on the coupling $g$. In general, when we increase $g$ we can expect to encounter an asymptotic critical value \cite{Arutyunov:2009ax} where some roots enter the physical strip, so that we must include appropriate driving terms. This would make the discussion more technically involved, but is of little relevance to understanding the lifting of degeneracies. We will therefore restrict our analysis to the small coupling region, below the first critical value of $g$.

For both states, it turns out that roots of $1+Y_{M|w}$, $1+Y_{M|vw}$ and $1-Y_-$, and poles of $Y_+$ play an important role. In order to discuss this, let us fix some notation. Roots related to state $\XX$ are described by script letters: $\RwXX{M}$ for $Y_{M|w}$, $\RwXX{0}$ for $Y_{-}$ and $\RvwXX{M}$ for $Y_{M|vw}$. They are fixed by the conditions
\begin{equation}
Y_{M|w}(\RwXX{M}-i/g)=-1,\ \ Y_{-}(\RwXX{0}-i/g)=1,\ \ Y_{M|w}(\RvwXX{M}-i/g)=-1\;.
\end{equation}
Similarly, we will use sans-serif letters to denote roots for state $\YY$: $\RwYY{M}$ for $Y_{M|w}$, $\RwYY{0}$ for $Y_{-}$  and $\RvwYY{M}$ for $Y_{M|vw}$, fixed by
\begin{equation}
Y_{M|w}(\RwYY{M}-i/g)=-1,\ \ Y_{-}(\RwYY{0}-i/g)=1,\ \ Y_{M|w}(\RvwYY{M}-i/g)=-1\;.
\end{equation}
In addition, for both states in both sectors $Y_+$ asymptotically has poles at the rapidities shifted by $i/g$, $Y_{+}(u_i-i/g)=\infty$, as in the Konishi case \cite{Arutyunov:2009ax}. The relevant roots are summarized in table \ref{tab:roots}.
\renewcommand{\arraystretch}{1.3}
\begin{center}
\begin{table}[!h]
\begin{center}
\begin{tabular}{| l | c | c | c | c | c | c | c | c |}
\hline
	& $1+\YLw{M}$ 			&  $1+\YRw{M}$ 			&  $1-\YLm$			 &$1-\YRm$			&$1+\YLvw{M}$ 		& $1+\YRvw{M}$ 	\\
\hline
{Roots $\XX$}	& $\pm \RwXX{M}-i/g$ 	& --					&$\pm \RwXX{0}-i/g$&--				& --				 & $\pm \RvwXX{M}-i/g$\\
{Roots $\YY$}	& $\RwYY{M}-i/g$ 		& $-\RwYY{M}-i/g$		&$- \RwYY{0}-i/g$	&$ \RwYY{0}-i/g$	& $ -\RvwYY{M}-i/g$	& $ \RvwYY{M}-i/g$\\
\hline
\end{tabular}
\end{center}
\caption{Roots for $Y$-functions in the left and right sectors for states $\XX$ and $\YY$ at small coupling. By definition we consider $\RwXX{M},\RwYY{M},\RvwXX{M},\RvwYY{M}>0$. Asymptotically, one observes that $\RwXX{M}\neq\RwYY{M}$ and $\RvwXX{M}\neq\RvwYY{M}$.}
\label{tab:roots}
\end{table}
\end{center}
\renewcommand{\arraystretch}{1}
We expect that the roots, and hence the driving terms, distribute differently between the left and right sectors for $\XX$ and $\YY$. This is indeed the case, as can be seen in table \ref{tab:roots}.

On the real mirror line, the asymptotic $Y$-functions for state $\XX$ are even, while for state $\YY$ the $Y$-functions do not have a definite parity but satisfy $Y^{(L)}(v) = Y^{(R)}(-v)$.

\subsubsection{The simplified TBA equations}
We now apply the contour deformation trick to the simplified TBA equations of \cite{Arutyunov:2009ur,Arutyunov:2009ux}. In order for the asymptotic solution to be a solution, we take the ground state TBA equations and define the integration contour such that it goes slightly below the line $-i/g$, \textit{i.e.} such that it encloses the poles of $Y_+$ at $u_i-i/g$ as well as the roots of table \ref{tab:roots} between itself and the real line. By taking the integration contour back to the real line, we find the appropriate driving terms, denoted $\driv$, and obtain the TBA equations. Below we list the driving terms that appear for each state. The integration kernels and $S$-matrices which enter in the equations below have been defined and are completely listed in \cite{Arutyunov:2009ax}. As usual, for any kernel or S-matrix we define $S^\pm(v):=S(v\pm i/g)$.
\vspace{10pt}\\

\bigskip
 \noindent
$\bullet$\ $M|w$-strings; $\ M\ge 1\ $, $Y_{0|w}=0$. The equation has the general form
\begin{equation}
\log Y^{(\alpha)}_{M|w} =  \log(1 +  Y^{(\alpha)}_{M-1|w})(1 +
Y^{(\alpha)}_{M+1|w})\star s
 + \delta_{M1}\, \log{1-{1\ov Y^{(\alpha)}_-}\ov 1-{1\ov Y^{(\alpha)}_+} }\hstar s +\driv^{(\alpha)}_{M|w},~~~~~
\end{equation}
where $\driv^{(\alpha)}_{M|w}$ are the driving terms that differ in the left and right sector for each given state and $\alpha =L,R$. For state $\XX$ we have
\begin{eqnarray}
\driv^{(L)}_{M|w}(v)&=&-\log S^{-}\!(\pm\RwXX{M-1}-v)-\log S^{-}\!(\pm\RwXX{M+1}-v)\;,\\
\driv^{(R)}_{M|w}(v)&=&0\;,
\nonumber
\end{eqnarray}
where the terms containing $\pm \varrho$ indicate the sum of two driving terms for opposite roots. For $\YY$ we have, instead,
\begin{eqnarray}
\driv^{(L)}_{M|w}(v)&=&-\log S^{-}\!(\RwYY{M-1}-v)-\log S^{-}\!(\RwYY{M+1}-v)\;,\\
\driv^{(R)}_{M|w}(v)&=&-\log S^{-}\!(-\RwYY{M-1}-v)-\log S^{-}\!(-\RwYY{M+1}-v)\;.
\nonumber
\end{eqnarray}
\vspace{10pt}\\
\bigskip
 \noindent
$\bullet$\ $M|vw$-strings; $\ M\ge 1\ $, $Y_{0|vw}=0$
\begin{align}
\log Y^{(\alpha)}_{M|vw}(v) = & - \log(1 +  Y_{M+1})\star s +
\log(1 +  Y^{(\alpha)}_{M-1|vw} )(1 +  Y^{(\alpha)}_{M+1|vw})\star
s\\
&  + \delta_{M1}  \log{1-Y^{(\alpha)}_-\ov 1-Y^{(\alpha)}_+}\hstar s+\driv_{M|vw}^{(0)} +\driv^{(\alpha)}_{M|vw}\,\nonumber.
\end{align}
When $M=1$, we find a driving term that is independent of state and sector, arising from the poles of $Y_+$, that is
\begin{equation}
\driv_{M|vw}^{(0)}=-\delta_{M1}\!\sum_{i=1}^4\log S^{-}\!(u_i-v)\;.
\end{equation}
In addition to that, for state $\XX$ we have
\begin{eqnarray}
\driv^{(L)}_{M|vw}(v)&=&0\;,\\
\driv^{(R)}_{M|vw}(v)&=&-\log S^{-}\!(\pm \RvwXX{M-1}-v)-\log S^{-}\!(\pm \RvwXX{M+1}-v)\;,
\nonumber
\end{eqnarray}
whereas for $\YY$ we have
\begin{eqnarray}
\driv^{(L)}_{M|vw}(v)&=&-\log S^{-}\!(- \RvwYY{M-1}-v)-\log S^{-}\!(- \RvwYY{M+1}-v),\\
\driv^{(R)}_{M|vw}(v)&=&-\log S^{-}\!(\RvwYY{M-1}-v)-\log S^{-}\!(\RvwYY{M+1}-v)\;.
\nonumber
\end{eqnarray}
\vspace{10pt}\\
\bigskip
 \noindent
$\bullet$\   $y$-particles
\begin{align}
\label{equ:tbaYproduct}
\log {Y^{(\alpha)}_+\ov Y^{(\alpha)}_-} = \, &  \log(1 +  Y_{Q})\star K_{Qy}+\driv^{(0)}_{ratio}\,,\\
\log {Y^{(\alpha)}_- Y^{(\alpha)}_+}= \, &  2 \log{1 +  Y^{(\alpha)}_{1|vw} \ov 1 +  Y^{(\alpha)}_{1|w} }\star s - \log\left(1+Y_Q \right)\star K_Q  \\
&+ 2 \log(1 +Y_{Q})\star K_{xv}^{Q1} \star s+\driv^{(0)}_{prod}+\driv^{(\alpha)}_{prod}\; \nonumber.
\end{align}
We expect both of these equations to pick up contributions from the exact Bethe equation\footnote{Here and afterwards, $*$ indicates analytic continuation to the string region. Also, $S_{1_*y}(u_{j},v) \equiv S_{1y}(u_{*j},v)$ is shorthand notation for the S-matrix with the first and second arguments in the string and mirror regions, respectively. The same convention is used for other kernels and S-matrices.} $Y_1(u_{*i}) = -1$, which will yield driving terms that do not depend on the state or sector. These are
\begin{eqnarray}
\driv^{(0)}_{ratio}(v)&=& -\sum_{i=1}^4 \log S_{1_*y}(u_i ,v)\,,\\
\driv^{(0)}_{prod}(v)&=& -\sum_{i=1}^4 \log {\big(S_{xv}^{1_*1}\big)^2\ov S_2}\star s( u_i,v)\,.\nonumber
\end{eqnarray}
where
\bea\nonumber
 &&\log  {\big(S_{xv}^{1_*1}\big)^2\ov S_2}\star s(u,v) \equiv  \int_{-\infty}^\infty\, dt\, \log  {S_{xv}^{1_*1}(u,t)^2\ov S_2(u-t)}\, s(t-v)
 \,.~~~~~
\eea
The contribution follows from the identity
\begin{equation}
\log{S_1(u_i-v)} - 2 \log{S_{xv}^{1_* 1}}\star s(u_i,v) = -\log{\frac{(S_{xv}^{1_* 1})^2}{S_2}}\star s(u_i,v) ,
\end{equation}
\noindent valid for real $u_i$.

\noindent In addition to the above driving terms, we have state-specific contributions in equation (\ref{equ:tbaYproduct}); for $\XX$ we have
\begin{eqnarray}
\driv^{(L)}_{prod}(v)&=& 2\log S^{-}\!(\pm \RwXX{1}-v)\,,\\
\driv^{(R)}_{prod}(v)&=&-2\log S^{-}\!(\pm \RvwXX{1}-v)\,,\nonumber
\end{eqnarray}
and for $\YY$ we have
\begin{eqnarray}
\driv^{(L)}_{prod}(v)&=&2\log \frac{S^{-}\!(\RwYY{1}-v)}{S^{-}\!(- \RvwYY{1}-v)}\,,\\
\driv^{(R)}_{prod}(v)&=&-2\log \frac{S^{-}\!(\RvwYY{1}-v)}{S^{-}\!(- \RwYY{1}-v)}\,.\nonumber
\end{eqnarray}
\vspace{10pt}\\
\bigskip
 \noindent
$\bullet$\ $Q$-particles
\begin{align}\label{equ:hybrid}
\log Y_Q(v) = & - L_{\scriptscriptstyle T\!B\!A}\, \tH_{Q} +\log \left(1+Y_{Q'} \right) \star \left(K_{\sl(2)}^{Q'Q}+2 s \star K^{Q'-1,Q}_{vwx}\right) +\driv^{(0)}_Q \\
&  +\!\!\!\sum_{\alpha\in\{L,R\}}\!\!\left(\log \(1 + Y^{(\alpha)}_{1|vw}\) \star s \hstar K_{yQ}+  \log \(1 + Y^{(\alpha)}_{Q-1|vw}\) \star s \phantom{\frac 1 1 }\right.\nonumber\\
  & \left. \phantom{+\sum_{\alpha\in\{L,R\}}}-   \log{1-Y^{(\alpha)}_-\ov 1-Y^{(\alpha)}_+} \hstar s \star K^{1Q}_{vwx}+ \frac{1}{2} \log {1- \frac{1}{Y^{(\alpha)}_-} \ov 1-\frac{1}{Y^{(\alpha)}_+} } \hstar K_{Q} \right.\nonumber\\
  & \left. \phantom{+\sum_{\alpha\in\{L,R\}}}+ \frac{1}{2} \log \big(1-\frac{1}{Y^{(\alpha)}_-}\big)\big( 1 - \frac{1}{Y^{(\alpha)}_+} \big) \hstar K_{yQ} +\driv^{(\alpha)}_Q \right).\nonumber
\end{align}
These are the TBA equations for $Q$-particles in the hybrid form of \cite{Arutyunov:2009ax}. Summation over repeated indices is understood. As before, we split the driving terms in a part independent of the specific state, that is $\driv^{(0)}_Q$, and sector dependent parts $\driv^{(\alpha)}_Q$ which will differ between $\XX$ to $\YY$. We then have
\begin{equation}
\driv^{(0)}_Q(v)=\sum_{i=1}^4\left(- \log S_{\sl(2)}^{1_*Q}(u_i,v)+ 2 \log{S}\star K^{1Q}_{vwx} (u_i,v)- \log{S^{1Q}_{vwx}} ( u_i,v)\right)\;,
\end{equation}
where for any kernel $K$ we define
\begin{equation}
 \log{S}\star K (u,v)=\lim_{\epsilon\to0^+}\int dt\; \log S\left(u-i/g-i\epsilon-t\right)\;K(t+i\epsilon,v)\;,
\end{equation}
which is the same type of contribution as for the Konishi state.

\noindent The left and right driving terms for $\XX$ are
\begin{eqnarray}\nonumber
\driv^{(L)}_Q(v)&=&\log S\star K^{1Q}_{vwx}(\pm\RwXX{0},v)-\frac{1}{2}\log S_Q^-(\pm\RwXX{0}-v)-\frac{1}{2}\log S_{yQ}(\pm\RwXX{0}-i/g,v),\\
\driv^{(R)}_Q(v)&=&-\log S\hstar K_{yQ}(\pm\RvwXX{1},v)-\log S^-(\pm\RvwXX{Q-1}-v),
\end{eqnarray}
while for $\YY$ we have
\begin{eqnarray}\nonumber
\driv^{(L)}_Q(v)&=&\log S\star K^{1Q}_{vwx}(\RwYY{0},v)-\frac{1}{2}\log S_Q^-(\RwYY{0}-v)-\frac{1}{2}\log S_{yQ}(\RwYY{0}-i/g,v)\\
& &-\log S\hstar K_{yQ}(-\RvwYY{1},v)-\log S^-(-\RvwYY{Q-1}-v),\\
\driv^{(R)}_Q(v)&=&\log S\star K^{1Q}_{vwx}(-\RwYY{0},v)-\frac{1}{2}\log S_Q^-(\RwYY{0}-v)-\frac{1}{2}\log S_{yQ}(-\RwYY{0}-i/g,v)\nonumber\\
& &-\log S\hstar K_{yQ}(\RvwYY{1},v)-\log S^-(\RvwYY{Q-1}-v).\nonumber
\end{eqnarray}
In the above, $K^{0,Q}_{vwx}=0$, $Y_{0|vw}=0$, meaning that the driving $\log S^+(v-\RvwYY{0})$ is not present.

Let us stress that in order to check (\ref{equ:hybrid}) on the asymptotic solution, $L_{\scriptscriptstyle T\!B\!A}$ needs to be specified. We find that
\begin{equation}
L_{\scriptscriptstyle T\!B\!A} = J+2\;,
\end{equation}
for both $\XX$ and $\YY$, just as for Konishi \cite{Arutyunov:2009ax}. As discussed in \cite{Arutyunov:2011uz}, $L_{\scriptscriptstyle T\!B\!A}$ is the maximal $J$ charge occurring in the conformal supermultiplet described by the TBA equations, and for a generic state that has full supersymmetry one indeed expects $L_{\scriptscriptstyle T\!B\!A}=J+2$. Nonetheless, there are examples of deformations of the superstring that break supersymmetry where different relations hold \cite{Arutyunov:2010gu,deLeeuw:2011rw}.

\subsubsection{The exact Bethe equations}
As discussed, the finite-size energies of states $\XX$ and $\YY$ depend on the allowed momenta. In the mirror TBA approach, these are found by analytically continuing the $Q$-particle TBA equations to the string region and imposing the exact Bethe equation $Y_1(u_{*i}) = -1$, which is the finite size quantization condition.

The (logarithm of the) exact Bethe equation  for a string rapidity $u_k$ is given by
\begin{align}\label{eq:ExactBethe}
(2n+1)\pi i = & i L_{\scriptscriptstyle T\!B\!A}\, p_k +\log \left(1+Y_{Q'} \right) \star \left(K_{\sl(2)}^{Q'1_*}+2 s \star K^{Q'-1,1_*}_{vwx}\right) +\driv^{(0)}_{1_*} \\
&  +\!\!\!\sum_{\alpha\in\{L,R\}}\!\!\left(\log \(1 + Y^{(\alpha)}_{1|vw}\) \star\left( s \hstar K_{y1_*}+ s^-\right)-   \log{1-Y^{(\alpha)}_-\ov 1-Y^{(\alpha)}_+} \hstar s \star K^{11_*}_{vwx} \right.\nonumber\\
  &\left.\ \ \ \ \ \ \ \ \ + \frac{1}{2} \log {1- \frac{1}{Y^{(\alpha)}_-} \ov 1-\frac{1}{Y^{(\alpha)}_+} } \hstar K_{1}+ \frac{1}{2} \log \big(1-\frac{1}{Y^{(\alpha)}_-}\big)\big( 1 - \frac{1}{Y^{(\alpha)}_+} \big) \hstar K_{y1_*} +\driv^{(\alpha)}_{1_*} \right),\nonumber
\end{align}
where the kernels have been analytically continued appropriately\footnote{See the appendix of  \cite{Arutyunov:2009ax} for details.}. As for the driving terms, we get the state independent contribution
\begin{eqnarray}
\driv^{(0)}_{1_*}(u_k)&=&\sum_{i=1}^4\left(- \log S_{\sl(2)}^{1_*1_*}(u_i,u_k)+ 2 \log{\rm Res}(S)\,\star K^{11_*}_{vwx} (u_i,u_k)\phantom{\frac 1 1}\right.\\
& &\phantom{\sum_{i=1}^4}\left.-  2 \log{(u_i - u_k - \tfrac{2i}{g})\,\frac{x_j^- -\tfrac{1}{x_k^-}}{x_j^- - x_k^+}}\right)\;.\nonumber
\end{eqnarray}
Coming to the state-dependent terms, for $\XX$ we have
\begin{eqnarray}\nonumber
\driv^{(L)}_Q(u_k)&=&\log S\star K^{11_*}_{vwx}(\pm\RwXX{0},u_k)-\frac{1}{2}\log S_1^-(\pm\RwXX{0}-u_k)-\frac{1}{2}\log S_{y1_*}(\pm\RwXX{0}-i/g,u_k),\\
\driv^{(R)}_Q(u_k)&=&-\log S\hstar K_{y1_*}(\pm\RvwXX{1},u_k)-\log S(\pm\RvwXX{1}-v),
\end{eqnarray}
while for $\YY$ we have
\begin{eqnarray}\nonumber
\driv^{(L)}_Q(u_k)&=&\log S\star K^{11_*}_{vwx}(\RwYY{0},u_k)-\frac{1}{2}\log S_1^-(\RwYY{0}-u_k)-\frac{1}{2}\log S_{y1_*}(\RwYY{0}-i/g,u_k),\\
& &-\log S\hstar K_{y1_*}(-\RvwYY{1},u_k)-\log S(-\RvwYY{1}-v),\\
\driv^{(R)}_Q(u_k)&=&\log S\star K^{11_*}_{vwx}(-\RwYY{0},u_k)-\frac{1}{2}\log S_1^-(-\RwYY{0}-u_k)-\frac{1}{2}\log S_{y1_*}(-\RwYY{0}-i/g,u_k),\nonumber\\
& &-\log S\hstar K_{y1_*}(\RvwYY{1},u_k)-\log S(\RvwYY{1}-v).\nonumber
\end{eqnarray}
We used the short-hand
\begin{equation}
\log {\rm Res}(S)\star K^{11_*}_{vwx} (u,v) = \int_{-\infty}^{+\infty}{\rm d}t\,\log\Big[S(u-i/g -t)(t-u)\Big] K_{vwx}^{11*}(t,v)\;,
\end{equation}
and indicated the momentum of the magnon as $p = i\tH_{Q}(z_{*})=-i\log{x_s(u+{i\ov g})\ov x_s(u-{i\ov g})}$.

Expanding the exact Bethe equation about the asymptotic $Y$-functions, we find, modulo $2\pi i$,
\begin{align}\label{equ:Rk}
\mathcal{R}_k \equiv &  \,2i\,p_k+\sum_{i=1}^4\left( 2 \log{\rm Res}(S)\,\star K^{11_*}_{vwx} (u_i,u_k)\phantom{\frac 1 1}-  2 \log{(u_i - u_k - \tfrac{2i}{g})\,\frac{x_j^- -\tfrac{1}{x_k^-}}{x_j^- - x_k^+}}\right)\\
&+\!\!\!\sum_{\alpha\in\{L,R\}}\!\!\left(- \log \mathscr{N}_{*}^{(\alpha)}+\log \(1 + Y^{(\alpha)}_{1|vw}\) \star\left( s \hstar K_{y1_*}+ s^-\right)-   \log{1-Y^{(\alpha)}_-\ov 1-Y^{(\alpha)}_+} \hstar s \star K^{11_*}_{vwx} \right.\nonumber\\
  &\left.\ \ \ \ \ \ \ \ \ \ \ \ + \frac{1}{2} \log {1- \frac{1}{Y^{(\alpha)}_-} \ov 1-\frac{1}{Y^{(\alpha)}_+} } \hstar K_{1}+ \frac{1}{2} \log \big(1-\frac{1}{Y^{(\alpha)}_-}\big)\big( 1 - \frac{1}{Y^{(\alpha)}_+} \big) \hstar K_{y1_*} +\driv^{(\alpha)}_{1_*} \right) = 0 \, , \nonumber
\end{align}
where the expression is evaluated at $u_k$\footnote{As in the Konishi case \cite{Arutyunov:2010gb}, this equation still holds for small perturbations around the solution of the Bethe-Yang equation, $\{u_k\}$.}. The terms $\log \mathscr{N}_{*}^{(\alpha)}$ arise from the analytic continuation of
\begin{equation}\label{equ:normalizfactor}
\mathscr{N}^{(\alpha)}(v)=\prod_{i=1}^{K^{\rm{II}}_{(\a)}}{{\frac{y^{(\a)}_i-x^-(v)}{y^{(\a)}_i-x^+(v)}\sqrt{\frac{x^+(v)}{x^-(v)}} \, }},
\end{equation}
that comes from the Bethe-Yang equations (\ref{equ:betheyangmain}), appearing whenever $K_{\a}^{\mathrm{II}}>0$. Equation (\ref{equ:Rk}) can be verified numerically to ensure that the analytic continuation has been performed correctly.
\smallskip

Since (\ref{equ:hybrid}) contains a sum over the left and right sectors, the form of the resulting exact Bethe equation is the same for $\XX$ and $\YY$. We might wonder whether this gives same momenta for both states, but this is of course not the case because the set of auxiliary $Y$-functions for the two states will be completely different. Indeed, even in the asymptotic case, the numerical value of the two set of roots is different: $\RwXX{M}\neq\RwYY{M}$ and $\RvwXX{M}\neq\RvwYY{M}$.
\bigskip

Finally, recall that the energy of each state is given by (\ref{eq:energy}). Since we have seen that the two set of TBA equations of $\XX$ and $\YY$ differ, we expect the energies $E^\XX$ and $E^\YY$ to be different as well. We will now show this explicitly by evaluating the first order wrapping corrections to the energy in both cases.

\subsection{Wrapping corrections}

As shown above, the TBA equations for the two states we consider are not equivalent. Therefore, the resulting $Y_Q$ functions and hence the energies should be different, thus lifting the degeneracy of the asymptotic Bethe ansatz. We will directly compute the leading order wrapping corrections to the energy to see this explicitly, naturally finding different results for the two states.

The leading order wrapping correction to the energy can be conceptually seen to arise from L\"uscher corrections \cite{Luscher:1985dn}, or equivalently by perturbatively expanding the free energy of the mirror model \cite{Arutyunov:2009ur}, depending on your point of view.

Using the asymptotic expression for the $Y_Q$-functions, (\ref{eq:YQasympt}), we can compute the leading order wrapping correction. To do so we evaluate our $Y_Q$-functions to lowest order in $g$, which give leading order wrapping interactions at seven loops. As expected the resulting $Y_Q$-functions are manifestly different. The expanded $Y_Q$-functions for either state are given in appendix \ref{ap:expansions}. Recall that the leading order wrapping correction to the energy is given by
\begin{align}
\nonumber
E_{LO} = -\frac{1}{2\pi}\sum_{Q=1}^{\infty}\int dv \frac{d\tilde{p}}{dv} Y^{\circ}_{Q}(v).
\end{align}
Integrating and summing the Y-functions for $J=4$ yields the following explicit wrapping correction for our states,
\begin{align}
\label{eq:wrappingJ4}
 E^\XX_{LO} & = \,  - \left(\tfrac{231}{32} \, \zeta(11) + \tfrac{21}{32} \, \zeta(9)- \tfrac{259}{32} \, \zeta(7)- \tfrac{113}{16} \, \zeta(5) +\tfrac{161}{32} \, \zeta(3) + \tfrac{1887}{1024} \right)\, g^{14}\\
  & \approx \, \, -0.2761\, g^{14},\nonumber\\
E^\YY_{LO} & = \, - \left(\tfrac{231}{32} \, \zeta(11) + \tfrac{105}{64}\, \zeta(9)- \tfrac{553}{64} \, \zeta(7)- \tfrac{589}{64} \, \zeta(5) +\tfrac{49}{8} \, \zeta(3) + \tfrac{2269}{1024} \right)\, g^{14} \\
& \approx  \, \, -0.1889\, g^{14}\nonumber.
\end{align}
This shows explicitly how the degeneracy present in the asymptotic Bethe ansatz is lifted by finite size (wrapping) corrections, with $\XX$ being the lighter state.

\section{Conclusion}

In this paper we have described a symmetry enhancement taking place for the $\AdS$ superstring in the asymptotic limit. Due to this enhancement certain states degenerate in the asymptotic limit, as described through the asymptotic Bethe ansatz. This symmetry is not present in the finite size model, indicating a \emph{qualitative} feature of the model that is not captured by the asymptotic solution. We illustrated these ideas on a set of two asymptotically degenerate states, by showing that they have manifestly different TBA equations, as well as explicitly computing their leading order wrapping corrections, clearly showing lifting of the asymptotic degeneracy. It would be interesting to verify these result on the gauge theory side, where two \emph{unrelated} sets of operators should have identical scaling dimensions exactly and only up to wrapping order.

\section*{Acknowledgments}

We are grateful to Gleb Arutyunov for useful discussions, and to Sergey Frolov, Marius de Leeuw and Ryo Suzuki for useful comments on the manuscript. The work by A.S. is part of the VICI grant 680-47-602 of the Netherlands Organization for Scientific Research (NWO). The work by S.T. is a part of the ERC Advanced Grant research programme No. 246974, {\it``Supersymmetry: a window to non-perturbative physics"}.

\appendix

\section{Transfer matrices and asymptotic $Y$-functions}
\label{ap:transfermatrices}
The eigenvalues of the transfer matrix $T^{(\alpha)}_{Q,1}$ in the $\sl(2)$-grading are known from \cite{Beisert:2006qh,Arutyunov:2009iq}. The index $\alpha=L,R$ labels the sector; for clarity we suppress it from $T_{Q,1}$ as well as from the auxiliary roots $y,w$ that parametrize the eigenvalues. We have
\begin{eqnarray}\label{eqn;FullEignvalue1}
&&T_{Q,1}(v)=\prod_{i=1}^{K^{\rm{II}}}{\textstyle{\frac{y_i-x^-}
{y_i-x^+}\sqrt{\frac{x^+}{x^-}} \, }}\left[1+
\prod_{i=1}^{K^{\rm{II}}}{\textstyle{
\frac{v-\nu_i+\frac{i}{g}Q}{v-\nu_i-\frac{i}{g}Q}}}\prod_{i=1}^{K^{\rm{I}}}
{\textstyle{\left[\frac{(x^--x^-_i)(1-x^-
x^+_i)}{(x^+-x^-_i)(1-x^+
x^+_i)}\frac{x^+}{x^-}  \right]}}  \right.\\
&&{\textstyle{+}}
\sum_{k=1}^{Q-1}\prod_{i=1}^{K^{\rm{II}}}{\textstyle{
\frac{v-\nu_i+\frac{i}{g}Q}{v-\nu_i+\frac{i}{g}(Q-2k)}}} \Big[
\prod_{i=1}^{K^{\rm{I}}}{\textstyle{\frac{x(v+(Q-2k)\frac{i}{g})-x_i^-}{x(v+(Q-2k)\frac{i}{g})-x_i^+}}}+
\prod_{i=1}^{K^{\rm{I}}}{\textstyle{\frac{1-x(v+(Q-2k)\frac{i}{g})x_i^-}{1-x(v+(Q-2k)\frac{i}{g})x_i^+}
}}\Big]\prod_{i=1}^{K^{\rm{I}}}{\textstyle{\frac{x^+-x_i^+}{x^+-x_i^-}\frac{v-v_i-(2k+1-Q)\frac{i}{g}}{v-v_i+(Q-1)\frac{i}{g}
}}}\nonumber\\
&& -\sum_{k=0}^{Q-1}\prod_{i=1}^{K^{\rm{II}}} {\textstyle{
\frac{v-\nu_i+\frac{i}{g}Q}{v-\nu_i+\frac{i}{g}(Q-2k)}}}\prod_{i=1}^
{K^{\rm{I}}}{\textstyle{\frac{x^+-x^+_i}{x^+-x^-_i}\sqrt{\frac{x^-_i}{x^
+_i}} \frac{v-v_i-(2k+1-Q)\frac{i}{g}}{v-v_i+(Q-1)\frac{i}{g}
}}}\prod_{i=1}^{K^{\rm{III}}}{\textstyle{\frac{w_i-v
+\frac{i(2k-1-Q)}{g}}{w_i-v+\frac{i(2k+1-Q)}{g}} }} \nonumber\\
&&\left. -\sum_{k=0}^{Q-1}\prod_{i=1}^{K^{\rm{II}}} {\textstyle{
\frac{v-\nu_i+\frac{i}{g}Q}{v-\nu_i+\frac{i
}{g}(Q-2k-2)}}}\prod_{i=1}^
{K^{\rm{I}}}{\textstyle{\frac{x^+-x^+_i}{x^+-x^-_i}\sqrt{\frac{x^-_i}{x^
+_i}} \frac{v-v_i-(2k+1-Q)\frac{i}{g}}{v-v_i+(Q-1)\frac{i}{g}
}}}\prod_{i=1}^{K^{\rm{III}}}{\textstyle{\frac{w_i-v+\frac{i}{g}(2k+3-Q)}{
w_i-v+\frac{i}{g}(2k+1-Q)}}}\right]. \nonumber
\end{eqnarray}
The variable
\begin{equation}
v=x^++\frac{1}{x^+}-\frac{i}{g}a=x^-+\frac{1}{x^-}+\frac{i}{g}a\,
\end{equation}
takes values in the mirror theory rapidity plane, so that $x^{\pm}=x(v\pm \frac{i}{g}a)$ where $x(v)$ is the mirror theory $x$-function.
Similarly, $x^{\pm}_j=x_s(u_j\pm \frac{i}{g})$, where $x_s$ is the string theory $x$-function. Recall that
\begin{align}
&x(u) = \frac{1}{2}(u-i\sqrt{4-u^2}), && x_s(u) = \frac{u}{2}(1+\sqrt{1-4/u^2}).
\end{align}
Notice that the transfer matrix comes with a prefactor of $\mathscr{N}=\prod_{i=1}^{K^{\rm{II}}}\frac{y_i-x^-}
{y_i-x^+}\sqrt{\frac{x^+}{x^-}}$ encountered already in the Bethe-Yang equations (\ref{equ:betheyangmain}) and in (\ref{equ:normalizfactor}). As discussed in \cite{Arutyunov:2011uz}, this is consistent with the requirement that $Y_{1*}(u_k)=-1$ on a solution of Bethe-Yang equations.
\smallskip

From the transfer matrix one can construct asymptotic $Y$-functions\footnote{The general construction of the Y-functions in terms of transfer matrices is based on the underlying symmetry group of the model \cite{Kuniba:1993cn,Tsuboi:2001ne}.  For the string sigma model asymptotic Y-functions were presented in \cite{Gromov:2009tv}. In fact, this solution can be directly derived from the Bajnok-Janik formula \cite{Bajnok:2008bm} and the AdS/CFT Y-system, see \cite{Arutyunov:2011uz}.} as follows
\begin{equation}
Y_{M|w}^{(\a)} = \frac{T_{1,M}^{(\a)} T_{1,M+2}^{(\a)}}{T_{2,M+1}^{(\a)}}  ,  \ \  Y_{-}^{(\a)} = -\frac{T_{2,1}^{(\a)}}{T_{1,2}^{(\a)}}  , \ \ Y_{+}^{(\a)} = - \frac{T_{2,3}^{(\a)}T_{2,1}^{(\a)}}{T_{1,2}^{(\a)}T_{3,2}^{(\a)}}, \ \ Y_{M|vw}^{(\a)} = \frac{T_{M,1}^{(\a)} T_{M+2,1}^{(\a)}}{T_{M+1,2}^{(\a)}},\label{eq:YmvwinT}
\end{equation}
in each sector, $\a=L,R$. Recall that the $Y_Q$ functions are given asymptotically by (\ref{eq:YQasympt}).

\section{Expansions for $Y^{\circ}_Q$ functions}
\label{ap:expansions}

Taking the transfer matrix (\ref{eqn;FullEignvalue1}) and expanding in the coupling constant yields the following expressions for the left and right transfer matrices of state $\XX$
\begin{align}
\label{eq:T2}
T^{(L)}_{Q,1}(\vec{u}|v)  = &\scriptstyle A_Q [6 Q^6+Q^4 \left(5 u_1^2+5 u_3^2+18 v^2+2\right)+Q^2 \left(u_1^4-2 v^2 \left(u_1^2+u_3^2+10\right)-2 \left(u_1^2+u_3^2+3\right)+u_3^4+18 v^4\right)\\
& \scriptstyle -v^4 \left(7 u_1^2+7 u_3^2+22\right)-\left(u_1^2+1\right) \left(u_3^2+1\right)\left(u_1^2+u_3^2+2\right)+v^2 \left(u_1^4+u_1^2 \left(8 u_3^2+6\right)+u_3^4+6u_3^2+2\right)+6 v^6] \nonumber\\
\label{eq:T0}
T^{(R)}_{Q,1}(\vec{u}|v)  = &\scriptstyle \frac{A_Q}{3}[ \left(u_1^2+u_3^2+2\right) \left(Q^4+Q^2 \left(3 u_1^2+3 u_3^2-2 v^2+2\right)+v^2 \left(3 u_1^2+3 u_3^2+2\right)-3 \left(u_1^2+1\right) \left(u_3^2+1\right)-3 v^4\right)]
\end{align}
where
\begin{equation}
\nonumber
A_Q  = {\textstyle \frac{8Q}{\left(u_1^2+1\right) \left(u_3^2+1\right) \left(Q^2+v^2\right)^2 \left(u_1^2+(Q-i v-1)^2\right) \left(u_3^2+(Q-i v-1)^2\right)}\,} .
\end{equation}
The expansion of the individual transfer matrices is rather convoluted for the $\YY$ state, so we present only the result for the product, given by
\begin{align}
\label{eq:T1T1}
T^{(L)}_{Q,1}T^{(R)}_{Q,1}(\vec{u}|v) = & \scriptstyle A_Q^2 (Q^2 +v^2) \left(u_1^2+u_3^2+2\right) \big[Q^8 \left(9 \left(u_1^2+u_3^2+2\right)-8 v^2\right)\\
&\scriptstyle \nonumber \quad +Q^6 \left(28 v^2 \left(u_1^2+u_3^2+2\right)+6
   \left(\left(u_1^2+u_3^2\right)^2-4\right)-32 v^4\right)\\
&\scriptstyle \nonumber  \quad +Q^4 \big(46 v^4 \left(u_1^2+u_3^2+2\right)+2 v^2 \left(u_1^4-6 u_1^2 \left(u_3^2+4\right)+u_3^4-24
   u_3^2-44\right)\\
&\scriptstyle \nonumber \quad  \quad \quad +\left(u_1^2+u_3^2+2\right) \left(u_1^4-2 u_1^2 \left(2 u_3^2+5\right)+u_3^4-10 u_3^2-2\right)-48 v^6\big)\\
&\scriptstyle \nonumber \quad +2 Q^2 \big(22 v^6 \left(u_1^2+u_3^2+2\right)-\left(u_1^2+1\right) \left(u_3^2+1\right) \left(\left(u_1^2+u_3^2\right)^2-4\right)\\
&\scriptstyle \nonumber \quad \quad \quad-v^4 \left(7 u_1^4+6 u_1^2 \left(5 u_3^2+8\right)+7 u_3^4+48 u_3^2+52\right)\\
&\scriptstyle \nonumber \quad \quad \quad+v^2 \left(u_1^2+u_3^2+2\right) \left(u_1^4+u_1^2 \left(8 u_3^2+2\right)+u_3^4+2 u_3^2+10\right)-16 v^8\big)\\
&\scriptstyle \nonumber \quad \quad \quad+17 v^8 \left(u_1^2+u_3^2+2\right)+\left(u_1^2+1\right)^2 \left(u_3^2+1\right)^2 \left(u_1^2+u_3^2+2\right)\\
&\scriptstyle \nonumber \quad \quad \quad-2 v^6 \left(5 u_1^4+6 u_1^2 \left(3 u_3^2+4\right)+5 u_3^4+24 u_3^2+20\right)\\
&\scriptstyle \nonumber \quad \quad \quad+v^4 \left(u_1^2+u_3^2+2\right) \left(u_1^4+2 u_1^2 \left(10 u_3^2+7\right)+u_3^4+14 u_3^2+22\right)\\
&\scriptstyle \nonumber \quad \quad \quad-2 \left(u_1^2+1\right) \left(u_3^2+1\right) v^2 \left(u_1^4+u_1^2 \left(6 u_3^2+4\right)+u_3^4+4 u_3^2\right)-8 v^{10} \big]
\end{align}
The S-matrix in the string-mirror region $S_{\sl(2)}^{1_*Q}$ is
found in \cite{Arutyunov:2009kf} (see also \cite{Bajnok:2009vm}) and has
the following leading behavior in $g$
\begin{equation}
S_{\sl(2)}^{1_*Q}(u,v)= {\textstyle-\frac{\big[(v-u)^2+(Q+1)^2\big]\big[Q-1 + i (v-u)\big]}{(u-i)^2 \big[Q-1-i( v-u)\big]} + O(g^2) }\, .
\end{equation}
These expressions are enough to build up the leading term in the weak-coupling expansion of the asymptotic function $Y^o_Q$, which is given by
\begin{align}
Y^{\circ}_Q(v) =\, \frac{g^{2J}}{Q^2+v^2} \frac{T^{(L)}(\vec{u}|v)T^{(R)}(\vec{u}|v)}{\prod_i S_0(v,u_i)}\, ,
\end{align}
where for our specific states we take either the product of (\ref{eq:T0}) and (\ref{eq:T2}), or (\ref{eq:T1T1}).

\end{document}